\documentclass[apjl]{emulateapj}
\usepackage{psfig,amsfonts,amsmath,graphicx,natbib,apjfonts}
\def\ra#1#2#3{#1$^{\rm h}$#2$^{\rm m}$#3$^{\rm s}$}
\def\dec#1#2#3{$#1^\circ#2'#3''$}
\def\nod{\nodata}
\def\grb{GRB\,070724A}

\def\har{1}
\def\ucb{2}
\def\psu{3}

\shorttitle{Near-IR Afterglow of Short GRB\,070724A}
\shortauthors{Berger et al.}

\begin{document}

\title{Discovery of the Very Red Near-Infrared and Optical 
Afterglow of the Short-Duration GRB 070724A}

\author{
E.~Berger\altaffilmark{\har},
S.~B.~Cenko\altaffilmark{\ucb},
D.~B.~Fox\altaffilmark{\psu}
and A.~Cucchiara\altaffilmark{\psu}
}

\altaffiltext{\har}{Harvard-Smithsonian Center for Astrophysics, 60
Garden Street, Cambridge, MA 02138}

\altaffiltext{\ucb}{Department of Astronomy, University of California,
Berkeley, CA 94720}

\altaffiltext{\psu}{Department of Astronomy and Astrophysics,
Pennsylvania State University, 525 Davey Laboratory, University Park,
PA 16802}

\begin{abstract} We report the discovery of the near-infrared and
optical afterglow of the short-duration gamma-ray burst GRB\,070724A.
The afterglow is detected in $iJHK_s$ observations starting 2.3 hr
after the burst with $K_s=19.59\pm 0.16$ mag and $i=23.79\pm 0.07$
mag, but is absent in images obtained 1.3 years later.  Fading is also
detected in the $K_s$-band between 2.8 and 3.7 hr at a $4\sigma$
significance level.  The optical/near-IR spectral index, $\beta_{\rm
O,NIR}\approx -2$, is much redder than expected in the standard
afterglow model, pointing to either significant dust extinction,
$A_V^{\rm host}\approx 2$ mag, or a non-afterglow origin for the
near-IR emission.  The case for extinction is supported by a shallow
optical to X-ray spectral index, consistent with the definition for
``dark bursts'', and a normal near-IR to X-ray spectral index.
Moreover, a comparison to the optical discovery magnitudes of all
short GRBs with optical afterglows indicates that the near-IR
counterpart of GRB\,070724A is one of the {\it brightest} to date,
while its observed optical emission is one of the faintest.  In the
context of a non-afterglow origin, the near-IR emission may be
dominated by a mini-supernova, leading to an estimated ejected mass of
$M\sim 10^{-4}$ M$_\odot$ and a radioactive energy release efficiency
of $f\sim 5\times 10^{-3}$ (for $v\sim 0.3$c).  However, the mini-SN
model predicts a spectral peak in the UV rather than near-IR,
suggesting that this is either not the correct interpretation or that
the mini-SN models need to be revised.  Finally, the afterglow
coincides with a star forming galaxy at $z=0.457$, previously
identified as the host based on its coincidence with the X-ray
afterglow position ($\sim 2''$ radius).  Our discovery of the
optical/near-IR afterglow makes this association secure, and
furthermore localizes the burst to the outskirts of the galaxy, with
an offset of $4.8\pm 0.1$ kpc relative to the host center.  At such a
large offset, the possible large extinction points to a dusty
environment local to the burst and rules out a halo or intergalactic
origin.  \end{abstract}


\section{Introduction}
\label{sec:intro}

The determination of sub-arcsecond positions for short-duration
gamma-ray bursts (GRBs) is of the utmost importance for our growing
understanding of their redshift distribution, energy scale, host
galaxies, and local environments.  Such localizations require the
detection of optical, near-infrared, or radio afterglows; or
alternatively an X-ray detection with the {\it Chandra} X-ray
Observatory.  As of August 2009, only 15 short GRBs have been
precisely localized in this manner, and all were detected in the
optical band\footnotemark\footnotetext{To date only two short GRBs
have been detected in the radio (050724 and 051221a;
\citealt{bpc+05,sbk+06}), and only one short GRB afterglow was
discovered with {\it Chandra} (050709; \citealt{ffp+05}).}
\citep{bpc+05,ffp+05,hwf+05,ltf+06,pcg+06,rvp+06,sbk+06,gcn6739,sap+07,gcn8190,pac+08,gcn8934,gcn8933,amc+09,gfl+09,gcn9342,pmg+09}.

Follow-up observations of some of these bursts have led to the
detection and characterization of host galaxies, most of them star
forming, and a small proportion ($\sim 20\%$) non-star forming (e.g.,
\citealt{ber09}).  The precise positions also allow us to study the
local environment of the bursts within their hosts, and current
observations point to offsets of $\sim 1-15$ kpc (e.g.,
\citealt{bpc+05,ffp+05,sbk+06,amc+09}).  However, since these are
projected positions, and since we generally lack detailed afterglow
observations that can shed light on the circumburst environment,
little is known about whether the bursts originate within the inner or
halo regions of their hosts (e.g., \citealt{sbk+06,lbb+09}).

Rapid optical observations have also been used to place limits on
emission from radioactive material synthesized in a putative
sub-relativistic outflow associated with a compact object binary
merger, a so-called Li-Paczynski mini-supernova \citep{lp98}.  Such
emission is theorized to have a typical peak time of $\sim 1$ d, and a
peak luminosity of $\sim 10^{42}$ erg s$^{-1}$, corresponding to
$m_{\rm AB}\sim 21$ mag at $z\sim 0.5$.  No such emission has been
detected to date (e.g., \citealt{hsg+05,bpp+06}).  Similarly, no
late-time emission from Type Ib/c supernova associations have been
detected (e.g., \citealt{hwf+05,bpp+06,sbk+06}).

Most recently, near-IR and optical non-detections of \grb\ have been
used to place limits on emission from a putative mini-SN associated
with this burst \citep{ktr+09}.  Here we report the detection of
near-IR and optical counterparts of \grb\ about 2.3 hr after the
burst, and show that the afterglow is actually one of the brightest
near-IR short GRB afterglows detected to date, but is one of the
faintest in the optical.  We use the observed fluxes and the unusually
red color to investigate the properties of the afterglow and/or
mini-SN, and to precisely measure the location of the burst relative
to its host galaxy.  Our discovery of the afterglow of \grb\ suggests
that recovery of a substantial fraction of short GRB optical/near-IR
afterglows requires observations to $m_{\rm AB}\sim 24$ mag within
about $0.5$ d.

\section{GRB\,070724A} 
\label{sec:obs}

\grb\ was discovered by the {\it Swift} satellite on 2007 July 24 at
10:53:50 UT with a duration of $0.40\pm 0.04$ s \citep{gcnr74}.  The
X-ray afterglow was detected with the on-board X-ray Telescope (XRT)
beginning 72 s after the burst, while no counterpart was detected with
the UV/Optical Telescope \citep{gcnr74}.  The X-ray position was
subsequently determined to a precision of $1.7''$ radius ($90\%$
containment).  An apparently extended source was detected in
coincidence with the XRT error circle in Digitized Sky Survey images,
and tentatively proposed as a possible host \citep{gcn6658}.
Subsequent near-IR and optical observations from UKIRT, Gemini-North,
the Palomar 60-inch telescope, and the VLT revealed that the source
was indeed an extended galaxy, but did not uncover an afterglow
\citep{gcn6662,gcn6664,gcn6666}.  A recent analysis using image
subtraction on UKIRT, NOT, CTIO 1.3-m, Keck, and VLT data similarly
reveals no afterglow to limits of $F_{\nu,K}\lesssim 30$ $\mu$Jy and
$F_{\nu,i}\lesssim 0.3$ $\mu$Jy at 3.2 and 22.2 hr after the burst,
respectively \citep{ktr+09}.

Spectroscopy of the galaxy within the XRT error circle revealed that
it is located at a redshift of $z=0.4571$, is undergoing active star
formation at a rate of 2.5 M$_\odot$ yr$^{-1}$, has a luminosity of
$L_B\approx 1.4$ L$^*$, and a metallicity of $12+{\rm log(O/H)}\approx
8.9$ (\citealt{ber09}; see also \citet{ktr+09} for similar results).
The resulting isotropic $\gamma$-ray energy in the observed $15-150$
keV range is $E_{\rm \gamma,iso}\approx 1.6\times 10^{49}$ erg.

\section{Discovery of the Near-IR and Optical Afterglow}
\label{sec:disc}

\begin{figure}
\epsscale{1}
\plotone{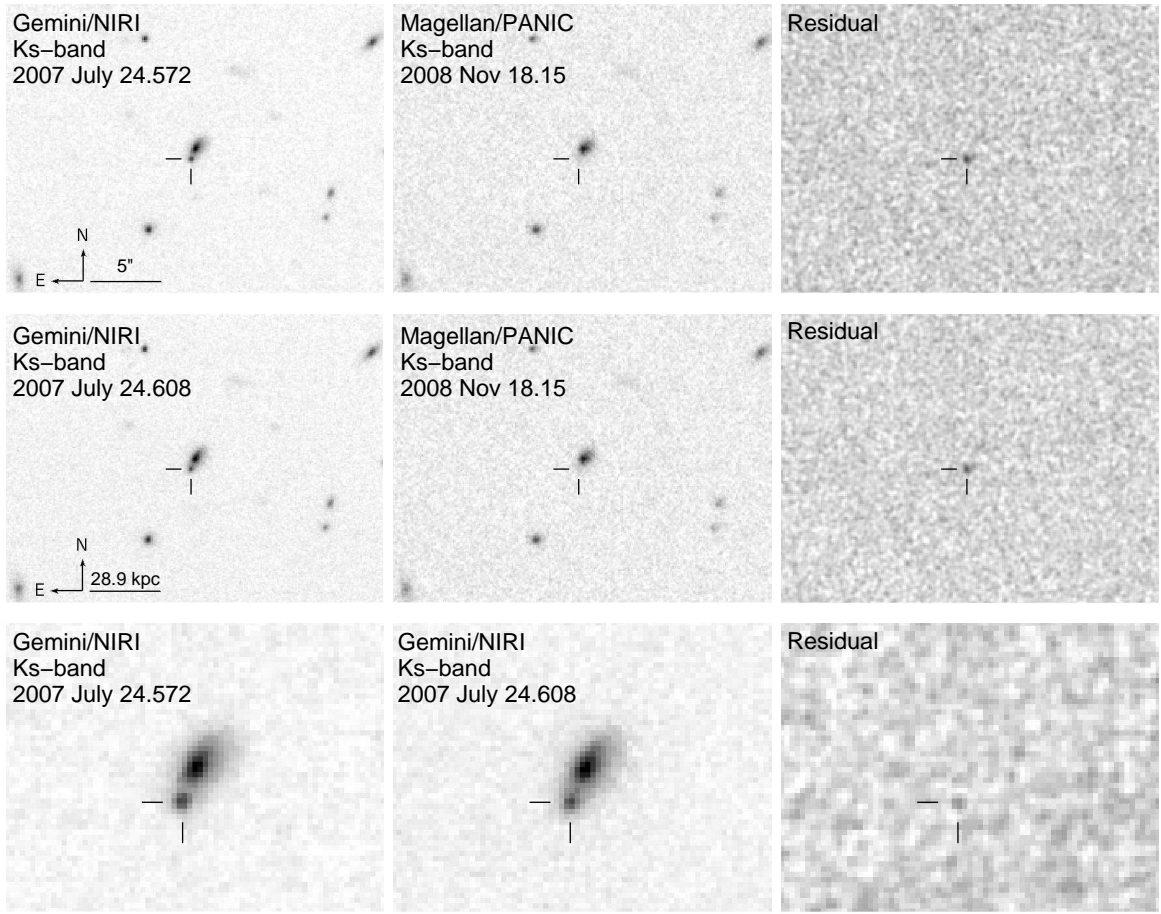}
\caption{Near-IR $K_s$-band images of \grb.  In each row we display
the afterglow image, corresponding template image, and residual image
from ISIS.  The near-IR afterglow is clearly visible in the residual
images relative to the final PANIC epoch, as well as in the
subtraction of the two NIRI epochs.
\label{fig:fig1}} 
\end{figure}

We observed the field centered on \grb\ with the Near Infra-Red Imager
and Spectrometer (NIRI) mounted on the Gemini-North 8-m telescope in
the $JHK_s$ bands starting on 2007 July 24.566 UT (2.69 hr after the
burst); see Table~\ref{tab:obs}.  The observations were obtained in
excellent seeing conditions, $\approx 0.35''$ in the $K_s$-band.  The
data were reduced using the {\tt gemini} package in IRAF, and
individual stacks were created in each filter.  Inspection of the
images reveals a point source coincident with the south-east edge of
the putative host galaxy; see Figure~\ref{fig:fig1}.

\begin{figure}
\epsscale{1}
\plotone{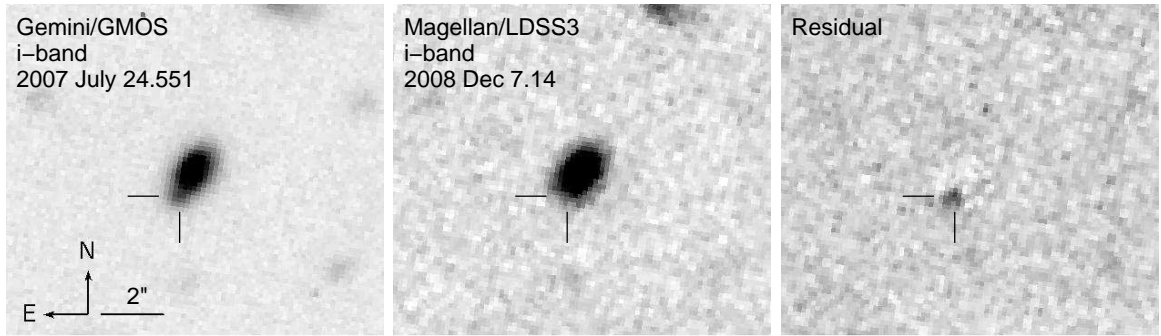}
\caption{Optical $i$-band images of \grb.  We display the afterglow
image, template image, and residual image from ISIS.  The afterglow is
clearly visible in the residual image.
\label{fig:fig4}} 
\end{figure}

Optical observations were obtained with the Gemini Multi-Object
Spectrograph (GMOS) mounted on the Gemini-North 8-m telescope in the
$gi$ bands starting on 2007 July 24.551 UT (2.33 hr after the burst;
Table~\ref{tab:obs}).  The data were reduced using the {\tt gemini}
package in IRAF.  The combined $i$-band image centered on the location
of \grb\ is shown in Figure~\ref{fig:fig4}.  An apparent extension is
seen in the same location as the near-IR source.

We obtained follow-up near-IR observations of the GRB with Persson's
Auxiliary Nasmyth Infrared Camera (PANIC) mounted on the
Magellan/Baade 6.5-m telescope on 2008 November 18.15 UT in the $K_s$
band (Table~\ref{tab:obs}).  The individual images were
dark-subtracted, flat-fielded, and corrected for bad pixels and cosmic
rays.  We then created object masks, which were used to construct
improved flat fields for a second round of reduction.  The data were
finally registered, shifted, and co-added.  The resulting combined
image is shown in Figure~\ref{fig:fig1}, and clearly reveals that the
point source visible in the NIRI images has faded away.

Similarly, late-time optical $gi$-band observations were obtained with
the Low Dispersion Survey Spectrograph (LDSS3) mounted on the
Magellan/Clay 6.5-m telescope on 2008 December 7.14 UT
(Table~\ref{tab:obs}).  The data were reduced using standard
procedures in IRAF.  The resulting $i$-band image is shown in
Figure~\ref{fig:fig4}.  As in the case of the late-time near-IR
observations, the extension seen in the early GMOS observations is no
longer detected.

To confirm the fading afterglow and to obtain accurate photometry and
astrometry we perform digital image subtraction on the NIRI and PANIC
$K_s$-band images and on the GMOS and LDSS3 $gi$ band images with the
ISIS package \citep{ala00}, which accounts for variations in the
stellar point-spread function (PSF).  We adopt the PANIC and LDSS3
images as templates with zero afterglow contribution since they were
obtained about 1.3 yr after the burst.  The resulting residual $K_s$-
and $i$-band images are shown in Figures \ref{fig:fig1} and
\ref{fig:fig4}, respectively, and clearly demonstrate that the point
source coincident with the host galaxy has faded away.  We therefore
conclude that this source is the afterglow of \grb.  We additionally
performed image subtraction on the two NIRI $K_s$-band observations
and find that the source has faded between mid-epochs of 2.832 and
3.696 hr after the burst at a $4\sigma$ confidence level
(Figure~\ref{fig:fig1}).  No residual is detected in the subtracted
$g$-band image.

\subsection{Absolute and Differential Astrometry}

We determine the absolute position of the afterglow from the NIRI and
GMOS residual images using the {\tt SExtractor} software
package\footnotemark\footnotetext{\tt
http://sextractor.sourceforge.net/}.  Since our NIRI images do not
contain any 2MASS stars, we first perform an astrometric tie of the
GMOS $i$-band image relative to USNO-B (using 13 common objects with a
resulting rms of $0.15''$) and then tie the NIRI astrometry to the
$i$-band image (using 20 common objects with a combined total rms of
$0.18''$).  The optical afterglow is located at
$\alpha$\,=\,\ra{01}{51}{14.071}, $\delta$\,=\,\dec{-18}{35}{39.33},
while the near-IR afterglow position is
$\alpha$\,=\,\ra{01}{51}{14.066}, $\delta$\,=\,\dec{-18}{35}{39.34}
(J2000).  These positions are consistent within the uncertainty of the
astrometric tie.  The optical/near-IR afterglow is offset by about
$0.5''$ and $0.9''$ relative to the X-ray positions from
\citet{ebp+09} and \citet{but07}, which have $90\%$ containment errors
of $1.6''$ and $1.7''$, respectively.

\begin{figure}
\epsscale{1}
\plotone{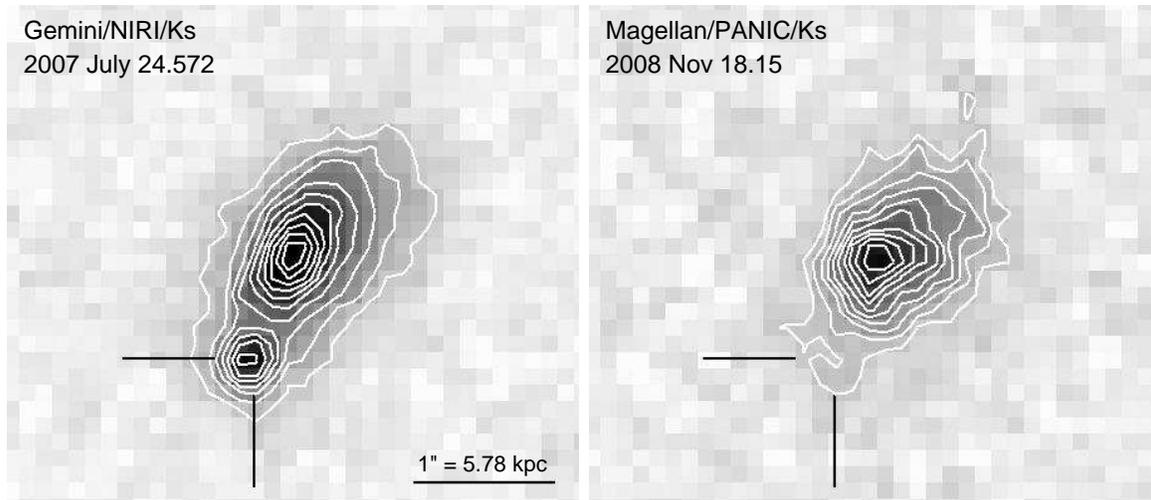}
\caption{Near-IR $K_s$-band images from NIRI and PANIC of the host and
afterglow of \grb.  The cross-hairs on the PANIC image mark the
location of the afterglow and indicates that \grb\ occurred on the
outskirts of the host galaxy.  The offset measured from the NIRI image
is $4.76\pm 0.06$ kpc.
\label{fig:fig2}} 
\end{figure}

The detection of the optical/near-IR afterglow makes the association
of \grb\ with the previously-proposed host galaxy secure
\citep{ber09}, and allows us to precisely measure the offset between
the GRB and host center.  We perform differential astrometry on the
NIRI images and find that the offset is $0.34\pm 0.01''$ east and
$0.75\pm 0.01''$ south of the host center, corresponding to a radial
offset of $0.82\pm 0.01''$.  The uncertainty reflects the centroiding
accuracy of both the afterglow and host, which we determine using {\tt
SExtractor}.  At a redshift of $z=0.4571$ the
scale\footnotemark\footnotetext{We use the standard cosmological
parameters, $H_0=71$ km s$^{-1}$ Mpc$^{-1}$, $\Omega_m=0.27$, and
$\Omega_\Lambda=0.73$.} is 5.785 kpc arcsec$^{-1}$, and the offset is
therefore $4.76\pm 0.06$ kpc.  This is similar to the offsets measured
for previous short GRBs with optical afterglows
\citep{bpc+05,ffp+05,sbk+06,amc+09}.  The location of the burst in the
late-time PANIC and LDSS3 images does not exhibit any excess emission;
see Figure~\ref{fig:fig2}.

\subsection{Photometry}

Photometry of the afterglow was performed on all residual images using
photometric standard stars that were observed in conjunction with the
PANIC and LDSS3 observations.  We find that the afterglow had
$K_s=19.59\pm 0.16$ mag and $K_s=19.64\pm 0.17$ mag in the first and
second NIRI epochs, respectively, and $i=23.79\pm 0.07$ mag in the
GMOS observation.  The $3\sigma$ limit on the $g$-band magnitude is
$g\gtrsim 23.5$ mag, determined by placing $\sim 10^3$ random
apertures on the residual image and using the width of the resulting
Gaussian flux distribution as $1\sigma$.  The near-IR magnitudes are
quoted in the Vega system, while the optical magnitudes are given in
the AB system.  Since the $g$-band limit is shallower than the
detected $i$-band magnitude, it provides no meaningful constraints on
the properties of the afterglow.

The observed $K_s$- and $i$-band magnitudes correspond to fluxes of
$9.3\pm 1.5$ $\mu$Jy, $8.9\pm 1.5$ $\mu$Jy, and $1.1\pm 0.1$ $\mu$Jy,
respectively.  We stress that the uncertainty in the flux of the
near-IR afterglow is dominated by the convolution with the PANIC
image, which was obtained under worse seeing conditions than the NIRI
images.  To assess the statistical uncertainty in the afterglow flux
we note that a stellar point source with $K_s=19.62$ mag, identical to
the afterglow brightness, located near the afterglow position has a
$1\sigma$ uncertainty of $0.03$ mag in the NIRI images.  This greater
depth in the NIRI data allows us to detect a significant fading
between the two NIRI observations despite an uncertainty of $0.16$ mag
relative to the PANIC observation.

\begin{figure}
\epsscale{1}
\plotone{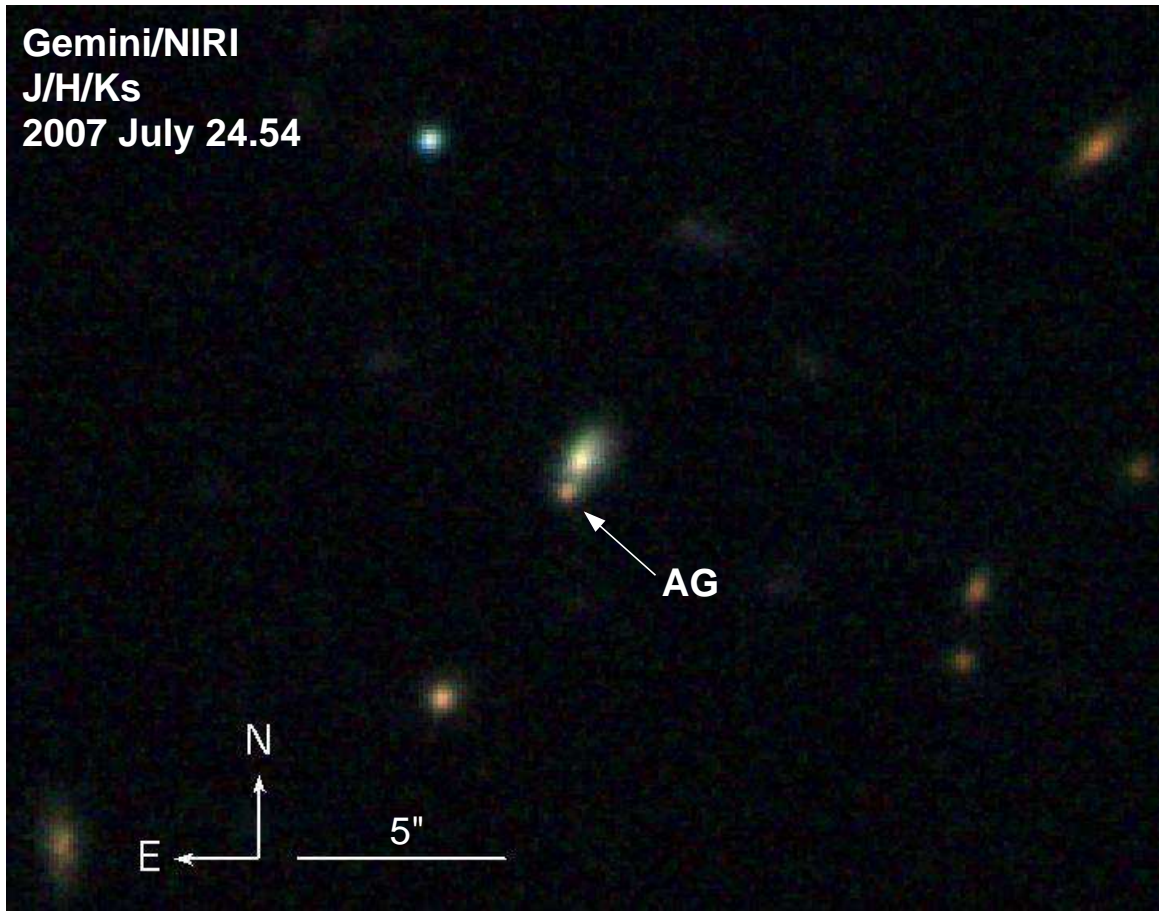}
\caption{Color composite image of the field of \grb\ using $J$ (blue),
$H$ (green) and $K_s$ (red) images from NIRI.  Although we cannot
clearly measure the afterglow flux in the $J$ and $H$ bands due to the
lack of late-time template images, we find that the afterglow is
clearly redder than the host galaxy.
\label{fig:fig3}} 
\end{figure}

We note that the afterglow is also clearly detected in the $J$- and
$H$-band images from NIRI.  However, due to the lack of late-time
template images we cannot robustly measure its brightness in these
bands.  Still, a color-composite image reveals that the afterglow is
redder in the near-IR bands than the rest of the host galaxy; see
Figure~\ref{fig:fig3}.

\section{Afterglow/Mini-Supernova Properties}
\label{sec:ag}

A comparison of our afterglow near-IR flux measurements to
contemporaneous limits from UKIRT observations by \citet{ktr+09}
reveals that the afterglow is about a factor of three times fainter
than the UKIRT upper limits.  These authors also find a limit on the
optical emission at 0.93 d after the burst of $F_{\nu,i}\lesssim 0.3$
$\mu$Jy.  A comparison to our detected $i$-band flux at 0.12 d
indicates that the afterglow temporal decay index is $\alpha<-0.6$
($F_\nu\propto t^\alpha$), typical of GRB afterglows.

On the other hand, a comparison of our contemporaneous $K_s$- and
$i$-band fluxes reveals an unusually steep spectral index,
$\beta=-2.0\pm 0.2$ ($F_\nu\propto \nu^\beta$).  Typically we expect
$\beta\approx -0.6$ to $-1.2$ for a wide range of electron power law
indices and values of the synchrotron cooling frequency \citep{spn98}.

The unusually red afterglow can be explained in two ways.  First, the
optical emission may be suppressed by extinction within the host
galaxy.  To reconcile the observed fluxes with a typical spectral
index of $\beta\approx -0.6$ requires $E(i-K_s)_{\rm obs}\approx 1$
mag, or a rest-frame $A_V^{\rm host}\approx 2$ mag for a Milky Way
extinction curve.  Such a large extinction seems unlikely given the
location of the afterglow at the edge of the host galaxy.  However, a
large average value of $E(B-V)\approx 1.2$ mag (i.e., $A_V\approx 4$
mag) has been inferred for the host galaxy based on its ratio of
H$\gamma$ and H$\beta$ emission lines \citep{ktr+09}, indicating that
extinction may indeed play a role in suppressing the optical emission.

We further investigate this possibility by comparison to the X-ray
afterglow brightness at the time of the optical observations,
$F_{\nu,X}(1\,{\rm keV})\approx 0.04$ $\mu$Jy \citep{gcnr74}.  This
leads to an optical to X-ray spectral index of $\beta_{\rm O,X}\approx
-0.5$, which marginally qualifies \grb\ as a ``dark burst''
\citep{jhf+04,ckh+09}.  On the other hand, the near-IR to X-ray
spectral index, $\beta_{\rm NIR,X}\approx -0.7$, is consistent with a
typical afterglow.  Thus, the comparison of the optical/near-IR and
X-ray afterglow emission is consistent with a standard afterglow
origin and significant dust extinction.  We note that \citet{ktr+09}
find excess absorption in the early X-ray data, but attribute this
result to rapid variations in the X-ray flux and spectral hardness.
In light of the possible significant dust extinction, the excess
photoelectric absorption may indeed be real.

An alternative explanation is that the near-IR flux is dominated by a
different source of emission than the afterglow.  In particular, in
the context of a compact object merger, the emission may be due to the
decay of radioactive material synthesized in a sub-relativistic
outflow, the so-called Li-Paczynski mini-supernova
(\citealt{lp98,rr02}; see also \citealt{ktr+09}).  In the formulation
of \citet{lp98}, the emission from such a mini-SN is described by a
peak luminosity ($L_p$):
\begin{equation}
L_p\approx 2\times 10^{44}\,f_{-3} M_{-2}^{1/2} (3\beta)^{1/2}
(\kappa/\kappa_e)^{-1/2}\,\,\,\,{\rm erg\,\,\, s^{-1}},
\label{eqn:lp}
\end{equation}
a peak effective temperature ($T_{{\rm eff},p}$):
\begin{equation}
T_{{\rm eff},p}\approx 2.5\times 10^4\,f_{-3}^{1/4} M_{-2}^{-1/8}
(3\beta)^{-1/8} (\kappa/\kappa_e)^{-3/8}\,\,\,\,{\rm K}, 
\label{eqn:Teff}
\end{equation}
and a peak time ($t_p$):
\begin{equation}
t_p\approx 1\,M_{-2}^{1/2} (3\beta)^{-1/2} (\kappa/\kappa_e)^{1/2}
\,\,\,\,{\rm d}
\label{eqn:tp}
\end{equation}
where $f$ is the fraction of rest mass energy released by the
radioactivity, $M$ is the ejecta mass in units of M$_\odot$,
$\beta\equiv v/c$ is the ejecta velocity, $\kappa$ is the average
opacity, $\kappa_e\approx 0.2$ cm$^2$ g$^{-1}$ is the electron
scattering opacity, and we use the notation $X\equiv 10^nX_{n}$.  

For our detected source we use the near-IR luminosity and observed
time as proxies for $L_p$ and $t_p$, respectively, leading to
$L_p\approx 10^{43}$ erg s$^{-1}$ and $t_p\approx 0.1$ d.  Using the
constraint that $3\beta\lesssim 1$ and assuming that $\kappa=
\kappa_e$, we find from Equation~\ref{eqn:tp} that $M_{-2}\lesssim
10^{-2}$ (i.e., $M\lesssim 10^{-4}$ M$_\odot$).  In conjunction with
Equation~\ref{eqn:lp} this provides a lower limit of $f_{-3}\gtrsim
5$.  The resulting lower limit on the effective temperature is
$T_{{\rm eff},p}\gtrsim 7\times 10^4$ K, corresponding to a peak in
the UV rather than in the near-IR.  The apparent discrepancy in the
spectral peak may be viewed as an indication that the observed
emission is not due to a mini-SN.  However, we note that
Equations~\ref{eqn:lp}-\ref{eqn:tp} correspond to the case of a power
law decay model with an assumed contribution from elements with a wide
range of decay timescales.  An exponential decay model, in which a
single element dominates the release of energy, may lead to distinctly
different luminosity and evolution \citep{lp98}.

To summarize, the unusually red optical/near-IR counterpart of \grb,
can be explained as a typical afterglow with significant dust
extinction, $A_V^{\rm host}\approx 2$ mag.  The alternative
explanation of a mini-SN leads to an expected peak in the UV, but this
may suggest that the mini-SN models should be revised.

\section{Discussion and Conclusions}

\begin{figure}
\epsscale{1}
\plotone{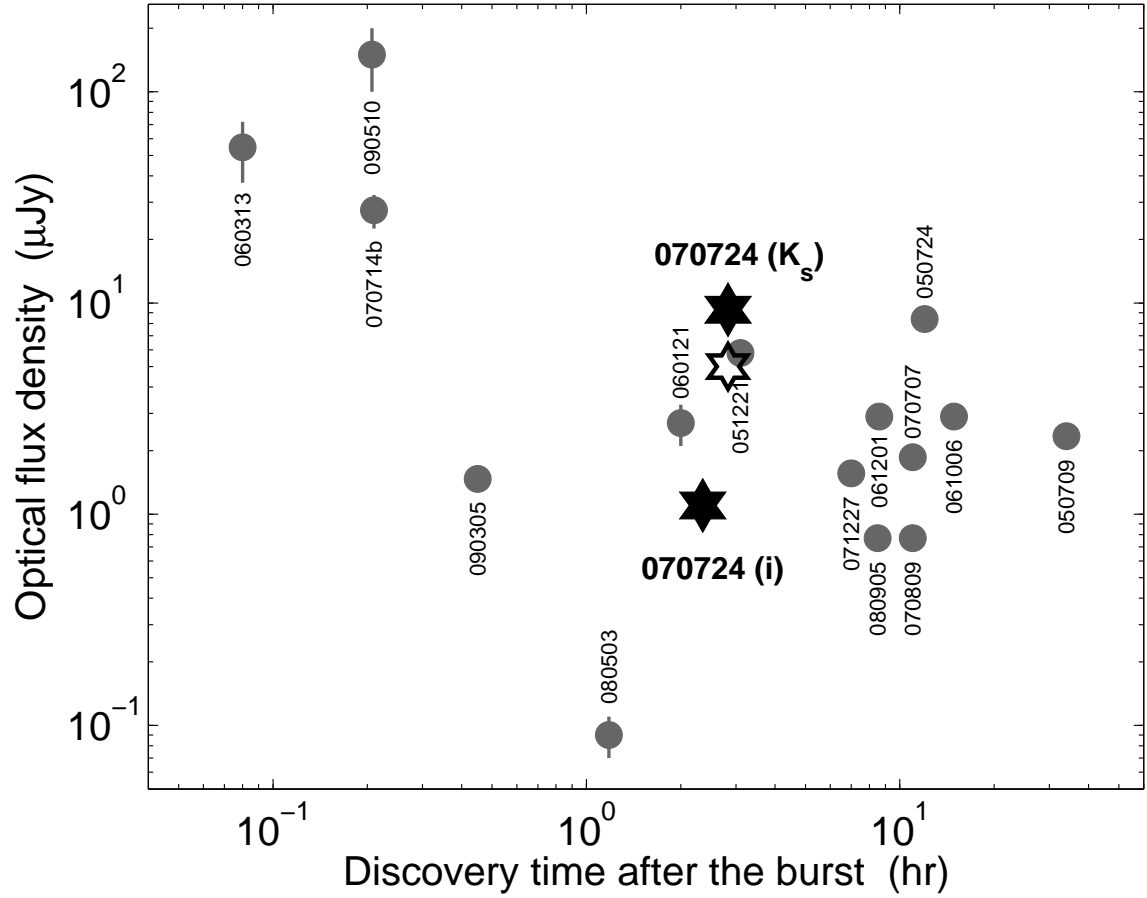}
\caption{Optical flux at the time of discovery for all 16 short GRBs
with optical and near-IR afterglows (including \grb).  The solid stars
mark the $i$- and $K_s$-band fluxes of the afterglow of \grb, while
the open star is the optical flux in the $i$-band extrapolated from
the near-IR with a spectrum of $F_\nu\propto\nu^{-0.6}$, typical of
GRB afterglows.  Data for other short GRBs are taken from the
literature \citep{hwf+05,bpc+05,sbk+06,pcg+06,ltf+06,rvp+06,amc+09,sap+07,pac+08,gfl+09,gcn6739,pmg+09,gcn8190,gcn8933,gcn8934,gcn9342}.
\label{fig:ag}} 
\end{figure}

Optical afterglow emission has now been detected from 16 short GRBs,
including \grb.  In Figure~\ref{fig:ag} we plot the flux of each
optical afterglow at the time of its discovery.  For \grb\ we show
both the optical and near-IR fluxes, as well as the expected $i$-band
flux extrapolated from the $K_s$-band using a typical spectral index
of $\beta=-0.6$.  While the observed $i$-band flux is one of the
faintest to date, the near-IR flux indicates that the afterglow of
\grb\ is actually one of the brightest at the time of its discovery.
Indeed, only the optical afterglows of GRBs 050724, 060313, 070714,
and 090510 were brighter, and of these only the afterglow of
GRB\,050724 was discovered on a comparable timescale; the optical
afterglows of GRBs 060313, 070714, and 090510 were all discovered
$\lesssim 20$ min after the burst.  On a timescale of 1 hr to 1 d
after the burst, the optical afterglows of short GRBs generally have
fluxes of $\sim 1-10$ $\mu$Jy, about two orders of magnitude lower
than the typical brightness of long GRB afterglows (e.g,
\citealt{kkz+08,ckh+09}).  From the existing distribution we conclude
that the detection of a substantial fraction of short GRB afterglows
requires optical/near-IR observations to $m_{\rm AB}\sim 24$ mag
within $\sim 0.5$ d.

The unusually red afterglow of \grb\ can be explained with a
substantial rest-frame dust extinction, $A_V^{\rm host}\approx 2$ mag.
This value is larger than the typical extinction inferred for most
long-duration GRBs, $A_V\sim 0.1-1$ mag (e.g., \citealt{pcb+09}), and
indeed the optical to X-ray spectral index, $\beta_{\rm OX}\approx
-0.5$, marginally qualifies \grb\ as a dark burst
\citep{jhf+04,ckh+09}.  Since the GRB is located on the edge of its
host galaxy, it is likely that the extinction arises in the local
environment of the burst.  This implies that the progenitor system was
not ejected from the host galaxy into the halo or intergalactic
medium.  Instead, the large extinction may point to an explosion site
within a star forming region, or alternatively that the progenitor
system itself produced the dust (for example, a binary system with an
evolved AGB star).  The possibility that some short GRBs are obscured
by dust has important ramifications for the nature of the progenitors,
and can also serve to localize the bursts to the galactic disk
environments.  Thus, rapid and deep near-IR observations are of
crucial importance.

Alternatively, in the context of a compact object merger model, the
near-IR emission may arise from radioactive decay in a
sub-relativistic outflow produced during the merger process -- a
mini-SN.  In this scenario, we find that the required ejected mass is
$M\lesssim 10^{-4}$ M$_\odot$, with a radioactive energy release
efficiency of $f\gtrsim 5\times 10^{-3}$.  We note, however, that in
the standard formulation the spectral peak at this time is expected to
be in the UV rather than in the near-IR.  This may indicate that the
detected source is completely due to afterglow emission, or that the
mini-SN models need to be revised.

\acknowledgements We thank Alicia Soderberg for assistance with the
digital image subtraction.  This paper includes data gathered with the
6.5 meter Magellan Telescopes located at Las Campanas Observatory,
Chile.  Observations were also obtained at the Gemini Observatory,
which is operated by the Association of Universities for Research in
Astronomy, Inc., under a cooperative agreement with the NSF on behalf
of the Gemini partnership: the National Science Foundation (United
States), the Particle Physics and Astronomy Research Council (United
Kingdom), the National Research Council (Canada), CONICYT (Chile), the
Australian Research Council (Australia), CNPq (Brazil) and CONICET
(Argentina)


\clearpage
\begin{deluxetable}{lcccccccc}
\tabletypesize{\footnotesize}
\tablecolumns{9}
\tabcolsep0.05in\footnotesize
\tablewidth{0pc}
\tablecaption{Log of Near-IR and Optical Observations of \grb\
\label{tab:obs}}
\tablehead {
\colhead {Date}             &
\colhead {$\Delta t$}       &
\colhead {Telescope}        &
\colhead {Instrument}       &
\colhead {Filter}           &
\colhead {Exposures}        &
\colhead {$\theta_{\rm FWHM}$} &
\colhead {mag}              &
\colhead {$F_\nu$}          \\
\colhead {(UT)}             &
\colhead {(d)}              &
\colhead {}                 &
\colhead {}                 &
\colhead {}                 &
\colhead {(s)}              &
\colhead {($''$)}           &
\colhead {}                 &
\colhead {($\mu$Jy)}           
}
\startdata
2007 July 24.549 & 0.094 & Gemini-N & GMOS  & $g$   & $2\times 180$ & $0.71$ & $\gtrsim 23.5$  & $\lesssim 1.5$ \\
2007 July 24.551 & 0.097 & Gemini-N & GMOS  & $i$   & $2\times 180$ & $0.53$ & $23.79\pm 0.07$ & $1.1\pm 0.1$ \\
2007 July 24.572 & 0.118 & Gemini-N & NIRI  & $K_s$ & $15\times 60$ & $0.35$ & $19.59\pm 0.16$ & $9.3\pm 1.5$ \\
2007 July 24.585 & 0.131 & Gemini-N & NIRI  & $J$   & $15\times 60$ & $0.45$ & \nod $^a$ & \nod $^a$ \\
2007 July 24.596 & 0.142 & Gemini-N & NIRI  & $H$   & $15\times 30$ & $0.46$ & \nod $^a$ & \nod $^a$ \\
2007 July 24.608 & 0.154 & Gemini-N & NIRI  & $K_s$ & $15\times 60$ & $0.35$ & $19.64\pm 0.17$ & $8.9\pm 1.5$ \\
2008 Nov 18.17   & 481.6 & Magellan & PANIC & $K_s$ & $54\times 20$ & $0.47$ & \nod $^b$ & \nod $^b$ \\
2008 Dec 7.13    & 500.6 & Magellan & LDSS3 & $g$   & $2\times 240$ & $0.94$ & \nod $^b$ & \nod $^b$ \\
2008 Dec 7.14    & 500.6 & Magellan & LDSS3 & $i$   & $3\times 120$ & $0.58$ & \nod $^b$ & \nod $^b$ 
\enddata
\tablecomments{$^a$ No templates are available in the $J$ and $H$
bands and as a result we cannot measure the flux of the afterglow.\\
$^b$ The flux of the afterglow in the PANIC and LDSS3 images is 
assumed to be zero.}
\end{deluxetable}

\clearpage

\clearpage

\clearpage

\clearpage

\end{document}